\documentclass[twocolumn,showpacs,preprintnumbers,amsmath,amssymb,aps]{revtex4}

\usepackage{graphicx}
\usepackage{dcolumn}
\usepackage{bm}

\begin{document}


\author{Peter B. Weichman}

\affiliation{BAE Systems, Advanced Information Technologies, 6 New
England Executive Place, Burlington, MA 01803}

\title{Second sound spectroscopy of a nonequilibrium
superfluid-normal interface}


\begin{abstract}

An experiment is proposed to test a previously developed theory of
the hydrodynamics of a nonequilibrium heat current-induced
superfluid-normal interface. It is shown that the interfacial
``trapped'' second sound mode predicted by the theory leads to a
sharp resonant dip in the reflected signal from an external second
sound pulse propagated towards the interface when its horizontal
phase speed matches that of the interface mode.  The influence of
the interface on thermal fluctuations in the bulk superfluid is
shown to lead to slow power dependence of the order parameter, and
other quantities, on distance from it.

\end{abstract}

\maketitle

\begin{figure}
\centerline{\includegraphics[angle=270,width=2.7in]{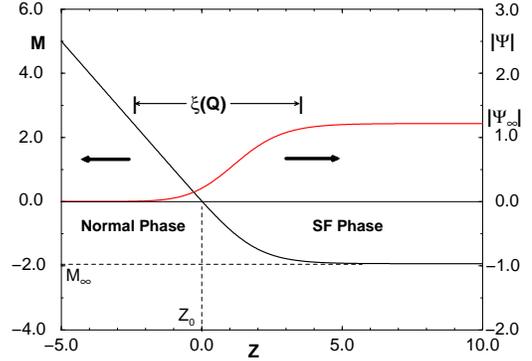}}

\caption{Scaled mean field steady state mean field temperature,
$M$, and order parameter magnitude, $|\Psi|$, profiles through the
interface as derived from (\ref{3}), using parameters $a_0=b_0=1$,
$d_0=2$ and $c_0=0$, which produce $M_\infty = -1.936$,
$|\Psi_\infty| = 1.216$ and $K_\infty = 0.676$.}

\label{fig:profile}
\end{figure}

In a seminal paper, Onuki \cite{Onuki} showed that when a uniform
heat current ${\bf Q} = Q {\bf \hat z}$ is driven through a sample
of $^4$He very close to the superfluid transition, a situation can
occur where the induced temperature profile divides the system
into separate upstream ($z < 0$) normal and downstream ($z > 0$)
superfluid regions. Within the interface between them (taken as
centered on $z=0$), the primary mode of heat transport converts
from thermal diffusion, with temperature gradient $\nabla T =
-{\bf Q}/\kappa$ where $\kappa$ is the thermal conductivity
(diverging at the superfluid transition), to superfluid
counterflow, with a supercurrent ${\bf j}_s$ flowing toward the
interface, and normal current ${\bf j}_n = -{\bf j}_s \propto {\bf
Q}$ carrying the heat away from it. The latter flow shorts out all
temperature gradients, leading to an asymptotically constant
temperature $T_\infty(Q)$ deep on the superfluid side, $z \to
\infty$ \cite{foot:vortex}. The interface width, $\xi(Q)$,
diverges as $Q \to 0$, and plays the role of the fundamental
correlation length in the system. Correspondingly, the order
parameter $\psi$ vanishes in the normal fluid, smoothly turns on
through the interface, and takes the helical form $\psi(z) =
|\psi_\infty(Q)| e^{-i k_\infty(Q) z}$ deep in the superfluid,
where the phase gradient $k_\infty = -m v_s/\hbar$ is proportional
to the superfluid velocity.  A mean field calculation of these
profiles is shown in Fig.\ \ref{fig:profile}.

\begin{figure}
\includegraphics[width=3.1in]{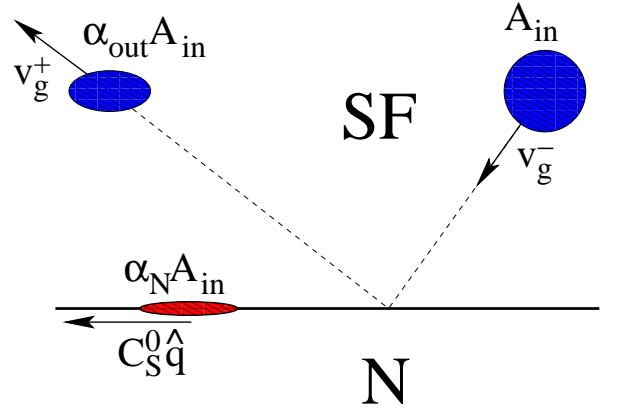}

\caption{Scattering geometry:  An incoming pulse with amplitude
$A_\mathrm{in}$ and spectrum narrowly centered on horizontal
wavevector ${\bf q}$ and frequency $\omega$ approaches the
interface at velocity ${\bf v}_g^-$. The reflected pulse, with
relative amplitude $\alpha_\mathrm{out}$, moves away at velocity
${\bf v}_g^+$, accompanied by an excitation of the interface
itself with relative amplitude $\alpha_N$ moving at speed $C_S^0$
along the direction of ${\bf q}$.  When the resonance condition
$\omega/q = C_S^0$ holds, $\alpha_\mathrm{out} \simeq 0$ (see
Fig.\ \ref{fig:reflect}) and the pulse is strongly absorbed by the
interface.}

\label{fig:scatt}
\end{figure}

This initial work spawned a series of experimental \cite{expt} and
theoretical \cite{HDtheory} investigations into this, and related,
nonequilibrium superfluid critical phenomena \cite{WHG}. Most
relevant to the present work, in Ref.\ \cite{WPMM} the
\emph{dynamics} of the interface under external forcing was
considered, and it was predicted that an interfacial second sound
mode exists, in which perturbations travel along the interface as
waves with a well defined sound speed $c(Q)$, and higher order
damping constant $D(Q)$. The waves are trapped in the sense that
their amplitude dies exponentially into the bulk phases on either
side \cite{WPMM}. In this paper an experiment is proposed, and the
underlying theory developed, to verify the existence of this mode
via second sound scattering. It is shown that when a pulse is
propagated toward the interface from the superfluid phase (see
Fig.\ \ref{fig:scatt}), strong resonant absorption occurs when its
horizontal phase speed matches $c(Q)$, leading to a sharp dip,
with depth scaling with $D(Q)$, in the reflected amplitude (Fig.\
\ref{fig:reflect} below). In a related effect, it is shown that
thermal excitation of these same modes leads to slow power law
corrections in $1/z$ to the order parameter and other quantities.

The analysis is based on the Model F equations of Halperin and
Hohenberg \cite{HalHo}.  These are derived from the Hamiltonian
\begin{eqnarray}
{\cal H} &=& \int d^dr \bigg\{\frac{1}{2} |\nabla \psi|^2
+ \frac{1}{2} r_0 |\psi|^2 + u_0 |\psi|^4
\nonumber \\
&+& \frac{1}{2} \chi_0^{-1}
(m-\chi_0 h_0 + \gamma_0 \chi_0 |\psi|^2)^2 \bigg\}
\label{1}
\end{eqnarray}
where $\psi({\bf r},t)$ is the superfluid order parameter, $m({\bf
r},t)$ is a linear combination of mass and energy density, and the
last term provides the crucial coupling between the two.  The
equations of motion are obtained from
\begin{eqnarray}
\partial_t \psi &=& -2\Gamma_0
\frac{\delta {\cal H}}{\delta \psi^*}
+ ig_0 \psi \frac{\delta {\cal H}}{\delta m} + \theta_\psi
\nonumber \\
\partial_t m &=& \lambda_0
\nabla^2 \frac{\delta {\cal H}}{\delta m}
- 2g_0 {\rm Im} \left(\psi^* \frac{\delta
{\cal H}}{\delta \psi^*} \right) + W + \theta_m.
\label{2}
\end{eqnarray}
where $\theta_\psi$ and $\theta_m$ are thermal white noise sources
and $W = Q[\delta(z-z_1) - \delta(z-z_2)]$ provides a source of
heat at $z_1 \to -\infty$ and a sink of heat at $z_2 \to +\infty$.
The local temperature is defined as $\mu({\bf r},t) = \delta {\cal
H}/\delta m({\bf r},t) = \chi_0^{-1} m + \gamma_0 |\psi|^2 - h_0$.
The various constant parameters may be partially determined by
fits to experimental data \cite{Onuki,HDtheory}.  The basic (mean
field) length in the problem is $l_0 = (\lambda_0/2 \chi_0
\gamma_0 Q)^{1/3}$, and it is convenient to define rescaled time
and space variables ${\bf R} ={\bf r}/l_0$ and $\tau = t/t_0$,
with $t_0 = l_0^2/\mathrm{Re} \Gamma_0$. The equations of motion
take the form
\begin{eqnarray}
&&\partial_\tau \Psi = -(1+ic_0)
[-\nabla_{\bf R}^2 + M + |\Psi|^2] \Psi
\nonumber \\
&&\ \ \ \ \ \ \ \ \
+\ ia_0 (M-M_\infty) \Psi + \Theta_\Psi
\nonumber \\
&&(b_0/2 a_0) \partial_\tau[d_0 M - |\Psi|^2]
= \nabla_{\bf R}^2 M + b_0 \nabla_{\bf R} \cdot {\bf J}
\nonumber \\
&&\ \ \ \ \ \ \ \ \
+\ \delta(Z-Z_1) - \delta(Z-Z_2) + \Theta_M,
\label{3}
\end{eqnarray}
where $a_0 = g_0/2\gamma_0\chi_0({\rm Re} \Gamma_0)$, $b_0 =
\gamma_0 \chi_0 g_0/2 \lambda_0 u_0$, $c_0 = {\rm Im}
\Gamma_0/{\rm Re} \Gamma_0$, $d_0 = 2u_0/\chi_0\gamma_0^2$, $\Psi
= 2l_0 \sqrt{u_0} e^{-ig_0 \tilde \mu t} \psi$, $M = [r_0 +
2\gamma_0\chi_0\mu]l_0^2$, $\tilde \mu = \lim_{z \to \infty}
\mu(z)$ is the asymptotic superfluid temperature, $M_\infty =
\lim_{Z \to \infty} M(Z)$, and ${\bf J} = {\rm Im}(\Psi^*
\nabla_{\bf R} \Psi)$ is the supercurrent density.  The rescaled
thermal noise sources $\Theta_\Psi = (2\sqrt{u_0}
l_0^3/\mathrm{Re} \Gamma_0) e^{-ig_0 \tilde \mu t} \theta_\psi$
and $\Theta_M = (2\gamma_0 \chi_0 l_0^4/\lambda_0) \theta_m$ have
covariances that diverge as $l_0^{4-d} \propto Q_0^{(d-4)/3}$.  An
expansion in $4-d$ is then required for a full theory in the small
$Q$ critical regime \cite{HDtheory}, but is difficult for the full
interface problem. However, the experiment proposed here is less
concerned with subtleties of nonequilibrium criticality, than with
basic hydrodynamical properties of the interface that are more
sharply defined and easier to investigate at larger $Q$.  A
correspondingly simpler theoretical approach will be taken in
which $Q$ is assumed large enough that the noise may be treated as
a perturbation on the homogeneous equations \cite{foot:scaling}.

Writing $\Psi = |\Psi| e^{i\phi}$ and ${\bf U} = [|\Psi|, \phi,
M]$, one may readily obtain a noise free steady state solution
with an interface (see Fig.\ \ref{fig:profile}), denoted by ${\bf
U}_0(Z)$, with $M_0(Z) = -Z$, $Z \to -\infty$, and ${\bf U}_0(Z)
\to [|\Psi_\infty|, -K_\infty Z, M_\infty]$, $Z \to +\infty$, with
the constraints $K_\infty^2 + |\Psi_\infty|^2 + M_\infty = 0$ and
$b_0 K_\infty |\Psi_\infty|^2 = 1$.

Long wavelength, harmonic excitations of the interface are
solutions of the form $\delta {\bf U} = {\bf U}({\bf R},\tau) -
{\bf U}_0 = \delta {\bf U}_{{\bf q},\omega}(Z) e^{i{\bf q} \cdot
(X,Y) - i \omega \tau}$, and may be thought of as being driven by
an external source on the plane $Z = Z_2$ with fixed frequency
$\omega$ and transverse wavevector ${\bf q}$.  These satisfy
linearized equations
\begin{equation}
i\omega {\bf \hat L}_3 \delta {\bf U}_{{\bf q},\omega}(Z)
= [{\bf \hat L}_0 + {\bf \hat L}_1 \partial_Z
+ {\bf \hat L}_2 (\partial_Z^2 - q^2)]
\delta {\bf U}_{{\bf q},\omega}(Z),
\label{4}
\end{equation}
with $Z$-dependent matrix coefficients which follow by
straightforward linearization of (\ref{3}) about ${\bf U}_0$, but
whose exact form will not be required here.
%
Deep on the superfluid side, the solution is a sum of incoming and
outgoing scattered waves,
\begin{equation}
\delta {\bf U}_{{\bf q},\omega}(Z)
= A_\mathrm{in} [{\bf V}^-_{{\bf q},\omega}
e^{i q_z^- Z} + \alpha_\mathrm{out}
{\bf V}^+_{{\bf q},\omega} e^{i q_z^+ Z}],
\label{6}
\end{equation}
obtained from the asymptotic form of (\ref{4}) with the
replacement $\partial_Z \to iq_z$. Here $A_\mathrm{in}$ is the
amplitude of the incoming excitation, and
$\alpha_\mathrm{out}({\bf q},\omega)$ is the relative amplitude of
the reflected wave. The wavevector components $q_z^+({\bf
q},\omega) > q_z^-({\bf q},\omega)$ are the two solutions to the
second sound dispersion relation
\begin{eqnarray}
0 &=& (1+d_0)\omega^2 + 4a_0 K_\infty q_z \omega
- 2a_0^2|\Psi_\infty|^2 q^2
\label{7} \\
&&-\ 2a_0^2 (|\Psi_\infty|^2 - 2K_\infty^2) q_z^2
+ O(\omega^3,q_z^3,q_z q^2,\ldots),
\nonumber
\end{eqnarray}
where stability requires that $|\Psi_\infty|^2 > 2K_\infty$
\cite{Onuki}. The corresponding eigenvectors are normalized to
have unit $\phi$-component, ${\bf V}^\pm_{{\bf q},\omega} =
[i(\omega + 2a_0 K_\infty q_z^\pm)/2a_0 |\Psi_\infty|, 1,
-i\omega/a_0]$ with $O[\omega^2,(q_z^\pm)^2,q^2,\ldots]$
corrections in the first and last components. To the exhibited
order, (\ref{7}) may be put in the elliptical form $(q_z -
q_{z,c})^2/(\Delta q_z)^2 + q^2/(\Delta q)^2 = 1$, with ellipse
center, and semi-major axes given by $q_{z,c} = K_\infty
\omega/a_0(|\Psi_\infty|^2 - 2K_\infty^2)$, $\Delta q_z = q_{z,c}
\sqrt{(1 + d_0) (|\Psi_\infty|^2/2K_\infty^2 - 1) + 1}$, $\Delta q
= \Delta q_z \sqrt{1 - 2K_\infty^2/|\Psi_\infty|^2}$. The
direction of pulse propagation is determined by the group velocity
\begin{equation}
{\bf v}_g({\bf q},\omega)
= \frac{\omega}{1 + q_{z,c}(q_z - q_{z,c})/\Delta q_z^2}
\left[\frac{\bf q}{(\Delta q)^2},
\frac{q_z - q_{z,c}}{(\Delta q_z)^2} \right],
\label{11}
\end{equation}
so that the $v_{g,z}^+ > 0 > v_{g,z}^-$ indeed have opposite sign,
and for each ${\bf q},\omega$ one may form pulses propagating
towards and away from the interface (see Fig.\ \ref{fig:scatt}).

The dissipative normal phase dynamics implies an exponentially
decaying asymptote $\delta M_{{\bf q},\omega} = A_\mathrm{in}
\alpha_N({\bf q},\omega) e^{i q_z^N Z}$, $Z \ll 0$, with
$\mathrm{Im} q_z^N < 0$ determined by the diffusion relation, $i
(b_0 d_0/2a_0) \omega = q^2 + (q_z^N)^2$. In solving (\ref{4}), we
will formally treat $\omega,q$ as small parameters of the same
order, thus viewing $C_S \equiv \omega/q$ as a fixed $O(1)$
parameter.  It follows that $q_z^\pm = q \sum_{k=0}^\infty
\pi_k^\pm q^{2k}$, $q_z^N = q^{1/2} \sum_{k=0}^\infty \pi_k^N q^k$
with known coefficients, $\pi_0^N = -(1+i) \sqrt{b_0 d_0
C_S/4a_0}$, $\pi_0^\pm = \pi_0^S \pm \Delta \pi_0^S \sqrt{1 -
1/(\Delta C)^2}$, where the ratios $\pi_0^S = q_{z,c}/q$, $\Delta
\pi_0^S = \Delta q_z/q$, $\Delta C = \Delta q/q$ are functions
only of $C_S$.

The appearance of $q^{1/2}$ in the normal phase leads one to
expect the solution to (\ref{4}) to have an expansion in $q^{1/2}$
rather than $q$ or $q^2$: $\delta {\bf U}_{{\bf q},\omega}(Z) =
\sum_{k=0}^\infty q^{k/2} \delta {\bf U}_k(Z)$,
$\alpha_{\mathrm{out},N} = \sum_{k=0}^\infty
\alpha_k^{\mathrm{out},N} q^{k/2}$. Asymptotic matching will
produce simultaneous expansions in powers of $q^{1/2}$ and $Z$
(coming from the expansion of the exponentials) on each side of
the interface, and matching each monomial coefficient will allow
determination of the unknown coefficients
$\alpha_k^{\mathrm{out},N}$. Defining ${\bf L}_Z = {\bf L}_0 +
{\bf L}_1 \partial_Z + {\bf L}_2
\partial_Z^2$, substitution of this expansion into (\ref{4}) leads to
the sequence of relations
\begin{eqnarray}
{\bf L}_Z \delta {\bf U}_0 &=& 0,\ {\bf L}_Z \delta {\bf U}_1 = 0
\nonumber \\
{\bf L}_Z \delta {\bf U}_2 &=& i C_S {\bf L}_3 \delta {\bf U}_0,\
{\bf L}_Z \delta {\bf U}_3 = i C_S {\bf L}_3 \delta {\bf U}_1
\nonumber \\
{\bf L}_Z \delta {\bf U}_4 &=& i C_S {\bf L}_3 \delta {\bf U}_2
- {\bf L}_2 \delta {\bf U}_0,
\label{15}
\end{eqnarray}
etc. Therefore, $\delta {\bf U}_{0,1}$ correspond to zero
frequency perturbations of the interface, i.e., infinitesimal
motions that simply produce a new steady state.  There are two of
these: a global phase rotation ${\bf v}_0 = [0,-K_\infty,0]$
(normalized using $K_\infty$ for later convenience), and a uniform
translation of the interface ${\bf v}_1 = \partial_Z {\bf U}_0$.
The asymptotes are ${\bf v}_1 \to {\bf v}_0$ for $Z \gg 0$,
$\delta M_1 \to -1$ for $Z \ll 0$. The appearance of ${\bf L}_Z
\delta {\bf U}_k$ in (\ref{15}) at each order means that the
solution can be determined only up to an arbitrary linear
combination $\mu_k {\bf v}_0 + \nu_k {\bf v}_1$ whose coefficients
must be determined from the matching conditions.

To begin, one obtains $\delta {\bf U}_{0,1} = \mu_{0,1} {\bf v}_0
+ \nu_{0,1} {\bf v}_1$. The $O(q^0,q^{1/2})$ matching conditions
yield $\nu_0 = -\alpha^N_0 = 0$, $\nu_1 = -\alpha^N_1$ on the
normal side, and $1 + \alpha^\mathrm{out}_0 = -K_\infty \mu_0$,
$\alpha^\mathrm{out}_1 = -K_\infty(\mu_1 + \nu_1)$ on the
superfluid side.

At $O(q)$ one obtains $\delta {\bf U}_2 = i \mu_0 C_s {\bf v}_2 +
\mu_2 {\bf v}_0 + \nu_2 {\bf v}_1$, in which ${\bf v}_2$ satisfies
${\bf L}_Z {\bf v}_2 = {\bf L}_3 {\bf v}_0$. It follows that ${\bf
v}_2$ is the change in shape of the interface under a perturbation
of the heat current, $Q \to 1 + \delta Q$, and results from
compression and rarefaction of the heat current in the vicinity of
the interface due to the incoming wave. Adjusting $Q$ leads to a
simple rescaling of ${\bf U}_0$, and one obtains exact result
${\bf v}_2 = (K_\infty/2 a_0 M_\infty) (Z {\bf v}_1 +
[|\Psi_0|,0,2M_0])$ \cite{WPMM}, with asymptotes ${\bf v}_2 \to
(K_\infty/2a_0 M_\infty) [|\Psi_\infty|, -K_\infty Z, 2M_\infty]$,
$Z \gg 0$, and $M_2 \to -3K_\infty Z/2a_0 M_\infty$, $Z \ll 0$.
The most important matching condition now arises from the term
linear in $Z$ on the superfluid side, which produces the
additional constraint $\alpha^\mathrm{out}_0 \pi_0^+ + \pi_0^- =
-\mu_0 C_S K_\infty^2/2a_0 M_\infty$, using which one obtains,
\begin{equation}
\alpha_0^\mathrm{out} = -(1 + K_\infty \mu_0)
= -\frac{\pi_0^- - K_\infty C_S/2a_0 M_\infty}
{\pi_0^+ - K_\infty C_S/2a_0 M_\infty},
\label{16}
\end{equation}
along with $\nu_2 = -\alpha^N_2$, and $\nu_1 = -\alpha^N_1 =
3K_\infty \mu_0 C_s/2 a_0 M_\infty \pi_0^N$. The latter
corresponds to the actual spatial amplitude of the interface
motion, $\delta Z({\bf q},\omega) = A_\mathrm{in} \nu_1 q^{1/2}$.
The numerator in (\ref{16}) vanishes, implying full absorption of
the incoming wave, for $C_S = C_{S,0}$, with
\begin{equation}
C_{S,0}^2 = \frac{4a_0^2 |\Psi_\infty|^2 M_\infty^2}
{2d_0 M_\infty^2 + |\Psi_\infty|^2
(2|\Psi_\infty|^2 - K_\infty^2)},
\label{17}
\end{equation}
corresponding precisely (the square of) the interfacial sound
speed (with physical value $c =  C_{S,0} l_0/\tau_0$) found in
Ref.\ \cite{WPMM}.

At order $O(q^{3/2})$ one obtains $\delta {\bf U}_3 = i C_s (\mu_1
{\bf v}_2 + \nu_1 {\bf v}_3) + \mu_3 {\bf v}_0 + \nu_3 {\bf v}_1$,
in which ${\bf v}_3$ satisfies ${\bf L}_Z {\bf v}_3 = {\bf L}_3
{\bf v}_1$.  It follows that ${\bf v}_3$ is the perturbation to
the interface profile induced by a change $\delta Q$ of the
incoming heat current on the normal side \emph{only}, with that
exiting on the superfluid side unchanged: ${\bf U}(Z,\tau) - {\bf
U}_0(Z-v\tau) \propto v {\bf v}_3(Z-v\tau)$. This causes heat to
build up behind the interface, moving it forward at an
instantaneous speed $v \propto \delta Q/Z_1$.  This heating effect
leads to dissipation of the interface motion, which is
\emph{singularly damped} \cite{WPMM}, at rate $\propto q^{3/2}$
rather than the bulk $q^2$ [which would arise from corrections to
(\ref{7})].

Although an analytic form for ${\bf v}_3$ is not available, it can
be obtained numerically.  For the purposes of determining $\mu_1,
\nu_2, \alpha_2^N, \alpha_1^\mathrm{out}$, only two numbers are
required, namely the coefficients $k_3,m_3$ in the asymptotic
forms $v_{3,\phi} \to k_3 Z$, $Z \gg 0$, and $v_{3,M} \to (b_0
d_0/4 a_0)Z^2 + m_3 Z$, $Z \ll 0$ (there may also be constant
terms, but these may be absorbed into $\mu_3,\nu_3$, which remain
undetermined at this order).  In addition to the $O(q)$ constraint
$\alpha_1^\mathrm{out} = -K_\infty(\mu_1+\nu_1)$, one obtains from
the matching:
\begin{eqnarray}
(\pi_0^N/C_S) \alpha_2^N + (3K_\infty/2a_0M_\infty) \mu_1
&=& \nu_1 m_3
\nonumber \\
(\pi_0^+/C_S) \alpha_1^\mathrm{out}
+ (K_\infty^2/2a_0M_\infty) \mu_1 &=& \nu_1 k_3,
\label{18}
\end{eqnarray}
with solution
\begin{eqnarray}
\alpha_1^\mathrm{out} &=&
-\frac{3C_S^2(\pi_0^+ - \pi_0^-)}
{2a_0 M_\infty \pi_0^N}
\frac{k_3 + K_\infty^2/2a_0 M_\infty}
{(\pi_0^+ - K_\infty C_S/2a_0 M_\infty)^2}
\nonumber \\
\alpha_2^N &=& -\nu_2 = \frac{C_S}{\pi_0^N} \left[m_3 \nu_1
+ \frac{3(\alpha_1^\mathrm{out}
+ K_\infty \nu_1)}{2a_0 M_\infty} \right].
\label{19}
\end{eqnarray}
Since $\pi_0^N$ is complex, the zero of the first order
combination $\alpha_\mathrm{out}(q) \approx \alpha_0^\mathrm{out}
+ q^{1/2} \alpha_1^\mathrm{out}$, is now shifted to the complex
value $\omega = C_{S,0} q + (1-i) D_{S,0} q^{3/2} + O(q^2)$, with
dissipation parameter
\begin{equation}
D_{S,0} = \frac{9 K_\infty (C_S^0)^{7/2}
(k_3 + K_\infty^2/2a_0 M_\infty)}
{4 \sqrt{a_0^3 b_0 d_0} M_\infty^2}.
\label{20}
\end{equation}
and corresponding physical value $D = D_S^0 l_0^{3/2}/\tau_0$.

Using the parameters $a_0 = b_0 = 1$, $d_0 = 2$, $c_0 = 0$, one
obtains $C_{S,0} = 1.089$, $k_3 = 0.488$, $D_{S,0} = 0.143$, and
$m_3 = 0.637$ \cite{WPMM}. The resulting magnitude and phase of
the reflection coefficient $\alpha_\mathrm(out)$, plotted as a
function of $C_S$ for various $q$, are shown in Fig.\
\ref{fig:reflect}.

Finally, consider the effects of thermal noise, under the
assumption, as discussed above, that the scaled noise amplitude is
sufficiently small that it may be treated within the linear
response regime.  Using the mode decomposition ${\bf U} =
\sum_{{\bf q},\omega} A_{{\bf q},\omega}(\tau) \delta {\bf
U}_{{\bf q},\omega}(Z) e^{i {\bf q} \cdot (X,Y)}$, one obtains an
equation of motion for the amplitude
\begin{equation}
(\partial_\tau + i\omega) A_{{\bf q}\omega}(\tau)
= \theta_{{\bf q},\omega}(\tau),
\label{21}
\end{equation}
in which the white noise $\theta_{{\bf q},\omega}(\tau)$ is the
appropriate projection of $\Theta_\Psi, \Theta_M$ onto the mode
eigenvector.  The solution
\begin{equation}
A_{{\bf q},\omega}(\tau) = \int_{-\infty}^\tau d\tau'
e^{-i\omega(\tau-\tau')} \theta_{{\bf q},\omega}(\tau'),
\label{22}
\end{equation}
(where the small dissipative negative imaginary part that gets
added to $\omega$ at order $q^2$ is actually required here to
ensure convergence) inserted back into ${\bf U}$, allows one to
compute stochastic averages of various quantities (calculations
are tedious and badly encumbered by matrix indices, and will be
presented in detail elsewhere). Not surprisingly, it is the phase
correlations that are the most important, decaying at long
distances as a power law $\langle \delta \phi({\bf R}) \delta
\phi({\bf R}') \rangle \propto |{\bf R}-{\bf R}'|^{2-d}$ (with
$d=3$ here) deep in the superfluid phase, with a complicated
anisotropic coefficient depending on the angle between ${\bf
R}-{\bf R}'$ and the heat flow direction ${\bf \hat z}$. The extra
$O(q)$ factors in the $|\Psi|,M$ components of ${\bf V}^\pm_{{\bf
q},\omega}$ produce weaker power laws $\langle \delta |\Psi({\bf
R})| \delta |\Psi({\bf R}')| \rangle, \langle \delta M({\bf R})
\delta M({\bf R}') \rangle \propto |{\bf R}-{\bf R}'|^{-d}$, also
with anisotropic coefficients.

\begin{figure}[tbp]

\includegraphics[width=3.2in]{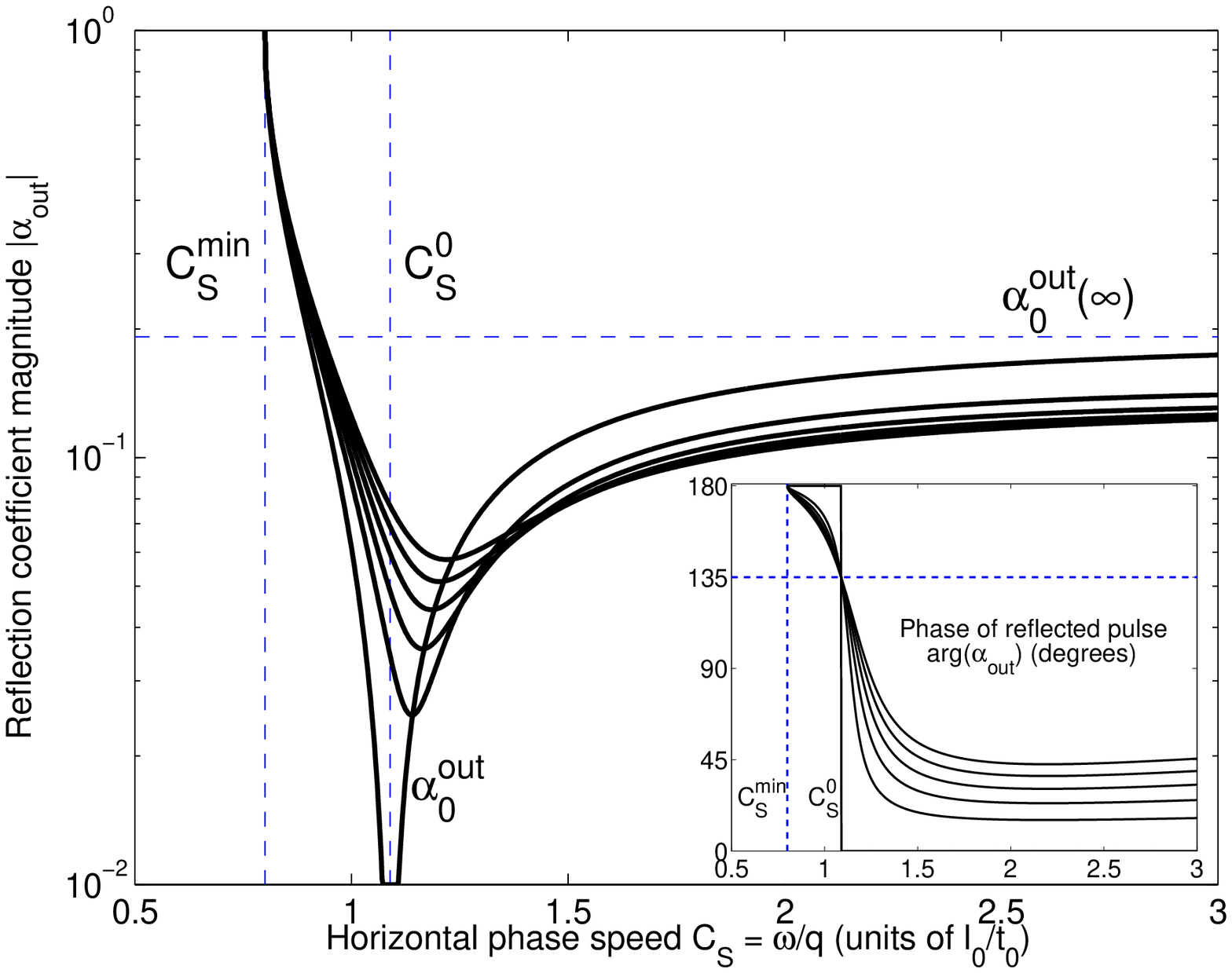}

\caption{Magnitude and phase (inset) of the reflection coefficient
$\alpha_\mathrm{out} = \alpha_0^\mathrm{out} +
\alpha_1^\mathrm{out} \sqrt{q}$ plotted as functions of the
horizontal phase speed $C_S$ for $q = 0,0.1,0.2,0.3,0.4,0.5$,
using scaled Model F parameters $a_0 = b_0 = 1$, $d_0 = 2$, $c_0 =
0$. In both figures, sharper curves correspond to smaller $q$. The
minimum allowed phase speed $C_S^\mathrm{min}$ occurs for $q_z^\pm
= q_{z,c}$, hence a purely horizontal group velocity. The large
$C_S$ asymptote of $\alpha_0^\mathrm{out}$ corresponds to a high
frequency pulse propagated nearly straight at the interface. For
small $q$ the minimum occurs at $C_S = C_{S,0} + D_{S,0} \sqrt{q}$
with value $B_0 D_{S,0} \sqrt{q}$, $B_0 = 2a_0^2 M_\infty^2
(|\Psi_\infty|^2 - 2K_\infty^2)/9K_\infty^2 |\Psi_\infty|^2
(C_S^0)^3$, scaling with the interfacial dissipation constant. The
phase switches rapidly (instantaneously for $q \to 0$) between
near-$\pi$ to near-zero as $C_S$ increases through $C_S^0$.}

\label{fig:reflect}
\end{figure}

Closer to the interface, there are residual correlations, coming
from interference between the incoming and reflected waves in
${\bf V}^\pm_{{\bf q},\omega}$, that produce $Z$-dependence in the
local fluctuations:  $\langle \delta \phi(\infty)^2 \rangle -
\langle \delta \phi({\bf R})^2 \rangle \propto 1/Z^{d-2}$,
$\langle \delta M(\infty)^2 \rangle - \langle \delta M({\bf R})^2
\rangle, \langle \delta |\Psi(\infty)|^2 - \delta |\Psi({\bf
R})|^2 \rangle \propto 1/Z^d$, where the argument $\infty$ is
shorthand for $Z \to \infty$. The phase fluctuations have a very
strong effect on the complex order parameter:
\begin{eqnarray}
\langle \Psi({\bf R}) \rangle
&\approx& \langle |\Psi({\bf R})| \rangle
\langle e^{i \phi({\bf R})} \rangle
= \Psi_0({\bf R}) e^{-\frac{1}{2}
\langle \delta \phi({\bf R})^2 \rangle}
\nonumber \\
&\approx& \Psi_0({\bf R})
e^{-\frac{1}{2} \langle \delta \phi(\infty)^2 \rangle}
(1 - B Z^{2-d}),
\label{23}
\end{eqnarray}
where $B$ is a coefficient, and the Gaussian property of the
noise has been used to average the exponential. There are two
interesting effects here: first, phase fluctuations can
significantly reduce the magnitude of the order parameter even in
the linear response regime [validity of (\ref{23}) requires only
the weaker assumption that the space-time derivatives of $\delta
\phi$, as opposed to $\delta \phi$ itself, be small]. Second, the
presence of the interface induces a slow power law approach of
$|\langle \Psi({\bf R}) \rangle|$ to its asymptotic superfluid
value, contrasting with the exponential approach of $\Psi_0({\bf
R})$.  This power law is not induced by the positional
fluctuations of the interface itself, which remain strongly
bounded \cite{WPMM}, but by the effects of its mere presence as a
reflecting boundary on the long-range bulk superfluid
correlations.

At linear order, the average temperature $\langle M({\bf R})
\rangle = M_0({\bf R})$ remains equal to its mean field value,
with the power law visible only in the variance.  However, it is
likely that similar power laws will be induced in $\langle M({\bf
R}) \rangle$ if nonlinear corrections are taken into account.


\begin{thebibliography}{99}
\bibitem{foot:vortex} Thermal noise actually induces the
occasional expanding vortex ring, giving rise to bulk dissipation,
and a hence a small temperature gradient, making $T_\infty(Q)$
less sharply defined, but this is a higher order effect that will
be ignored in the present work.
\bibitem{foot:scaling} Scaling arguments, similar to those used
in Ref.\ \cite{WPMM}, may be used to extrapolate the phenomenology
into the critical regime.
\end{thebibliography}
\end{document}